# Atomic and electronic structure of two-dimensional Mo$_{(1-x)}$W$_x$S$_2$ alloys

**Xue Xia[1§], Siow Mean Loh[1§], Jacob Viner[2], Natalie C. Teutsch[1], Abigail J. Graham[1], Viktor Kandyba[3], Alexei Barinov[3], Ana M. Sanchez[1], David C. Smith[2], Nicholas D. M. Hine[1*], Neil R. Wilson[1*]**

[1] Department of Physics, University of Warwick, Coventry, CV4 7AL, UK
[2] School of Physics and Astronomy, University of Southampton, Southampton SO17 1BJ, UK
[3] Elettra - Sincrotrone Trieste, S.C.p.A., Basovizza (TS), 34149, Italy

§ these authors contributed equally

E-mail: neil.wilson@warwick.ac.uk, n.d.m.hine@warwick.ac.uk

## Abstract

Alloying enables engineering of the electronic structure of semiconductors for optoelectronic applications. Due to their similar lattice parameters, the two-dimensional semiconducting transition metal dichalcogenides of the MoWSeS group (MX$_2$ where M= Mo or W and X=S or Se) can be grown as high-quality materials with low defect concentrations. Here we investigate the atomic and electronic structure of Mo$_{(1-x)}$W$_x$S$_2$ alloys using a combination of high-resolution experimental techniques and simulations. Analysis of the Mo and W atomic positions in these alloys, grown by chemical vapour transport, shows that they are randomly distributed, consistent with Monte Carlo simulations that use interaction energies determined from first-principles calculations. Electronic structure parameters are directly determined from angle resolved photoemission spectroscopy measurements. These show that the spin-orbit splitting at the valence band edge increases linearly with W content from MoS$_2$ to WS$_2$, in agreement with linear-scaling density functional theory (LS-DFT) predictions. The spin-orbit splitting at the conduction band edge is predicted to reduce to zero at intermediate compositions. Despite this, polarisation-resolved photoluminescence spectra on monolayer Mo$_{0.5}$W$_{0.5}$S$_2$ show significant circular dichroism, indicating that spin-valley locking is retained. These results demonstrate that alloying is an important tool for controlling the electronic structure of MX$_2$ for spintronic and valleytronic applications.

## 1. Introduction

Semiconducting transition metal dichalcogenide monolayers, such as the MoWSeS group (MX$_2$ where M= Mo or W and X=S or Se), are of great interest both for their practical applications and for the fundamental science that can be studied in them.[1] Much of the interest has focused on the optical properties of MX$_2$: they exhibit a transition from indirect gap in bulk to direct gap in monolayers, and strong Coulomb interactions lead to high exciton binding energies[2]. In the monolayers, spin-orbit coupling (SOC) splits both the valence band maximum (VBM) and conduction band minimum (CBM) to give spin-polarised bands at the Brillouin zone corners.[3,4] This leads to spin-valley locking, [5–8] with the potential for optically generated spin-polarised currents, and the resultant interest in valleytronics[9] and spintronics[10] in these materials.

Band engineering through alloying has been essential to the development of III-V semiconductors, giving control over band parameters such as band gaps and band alignments, leading to their application in commercial optoelectronics. Following this, theoretical and experimental works have shown that for $MX_2$ alloying allows not only continuous tuning of the band gap over a large range, from 1.55 to 1.99 eV,[11–18] but also control over the magnitude of SOC in both the conduction and valence band.[13,14,19] Combined, these band parameters have dramatic influences on the optical and electrical properties of $MX_2$. However, due to the difficulty in accurately determining the single-particle electronic structure, there is still uncertainty over how the electronic structure changes with alloying. The high exciton binding energies complicate the determination of electronic structure parameters from conventional optical spectroscopy measurements, such as photoluminescence (PL). This leaves some outstanding questions: for example the SOC is predicted to change linearly with composition, but inferring the SOC from the energy difference between A and B excitons suggested a non-linear bowing in $Mo_{(1-x)}W_xSe_2$ monolayers.[20]

With the comparatively small difference in lattice parameters, high-quality MoWSeS alloys are relatively easy to synthesize, either through direct growth of monolayers by chemical or physical vapour deposition[11,17,19,21–23], or by mechanical exfoliation of bulk alloy crystals grown by chemical vapour transport.[14,24–26] In such alloys, the local atomic arrangement can play an important role in determining the resultant properties of the material. Both atomically random[14,19,25] and ordered[27–29] alloying have previously been observed in transition metal dichalcogenide monolayers and theoretical studies have suggested the possibility of order-disorder phase transitions.[30,31] Study of the ordered alloy phases revealed sensitivity of the bandstructure at both the VBM and the CBM to composition, but specifically also sensitivity at the CBM to ordering.[31]

In this report we study exfoliated flakes of CVT grown $Mo_{1-x}W_xS_2$ alloys. By comparing quantitative analysis of the atomic structure determined by scanning transmission electron microscopy (STEM) to Monte Carlo simulations, we reveal that the atomic arrangements are consistent with those expected from thermodynamic considerations. Starting from these atomic structure models, and using large unit cells to avoid simulation artefacts, we apply linear scaling DFT to predict electronic structure changes with alloying. Comparison with angle-resolved photoemission spectroscopy (ARPES) valence band measurements shows that the SOC at the VBM does indeed scale linearly with stoichiometry ($x$). The predictions indicate that at intermediate compositions the SOC at the CBM should be less than the disorder potential in the alloys. Despite this, polarisation-resolved photoluminescence measurements of a $Mo_{0.5}W_{0.5}S_2$ monolayer show that spin-valley locking is retained and hence that such alloys have a promising future in spintronic applications.

## 2. Methods

*Crystal growth and exfoliation.* Single crystals were synthesised by chemical vapor transport (CVT) in a two-step process.[32] First, Mo (purity 99.9%, Sigma-Aldrich), W (purity 99.9%, Sigma-Aldrich) and S (purity 99.9%, Sigma-Aldrich) element powders were mixed stoichiometrically into an ampoule. The ampoule was pumped down to a pressure of $10^{-6}$ mbar and sealed. The mixture was heated to 1000 °C for 3 days to form $Mo_{1-x}W_xS_2$ powder. In the second step, crystals were grown from this powder. The synthesised compounds were transferred to a new quartz ampoule with larger diameter and mixed with the transport agent, $I_2$ (10 mg cm$^{-3}$ of the ampoule volume). To keep the $I_2$ stable, the ampoule was evacuated to $10^{-6}$ mbar in ice and sealed. The ampoule was placed into a three-zone furnace as shown in Supplementary Material section 1 (SuppMat S1), the charge zone was kept at 1050 °C for 20 days with the growth zone at 950 °C. After cooling to room temperature, the ampoule was opened in air and single crystals were collected at the growth end as shown in the schematic figure in SuppMat S1.

For PL measurements, the as-grown crystals were mechanically exfoliated onto a Si/SiO$_2$ wafer using scotch tape. Prior to peeling off, the substrates were heated for a few minutes on a hot plate at 130°C.[33] Monolayer flakes were identified by their optical contrast, by their Raman spectra, and by their PL emission. PL and Raman spectra were acquired at room temperature with a confocal microspectrometer (Renishaw inVia Reflex Raman microscopes) with 442 nm and 532 nm excitation.

*Compositional analysis.* Crystal composition was analysed by a combination of X-ray photoelectron spectroscopy (XPS) and energy dispersive X-ray analysis (EDX), further details are given in SuppMat S2. EDX showed uniform composition across individual crystals, indicating homogeneity at the microscale. The compositions determined by XPS and EDX were consistent with each other and with those measured from analysis of the atomic resolution STEM images, indicating homogeneity also at the nanoscale.

*Atomic resolution imaging.* Flakes were mechanically exfoliated onto chemical vapour deposition (CVD) grown graphene on copper [34], then a layer of CVD graphene was wet transferred on top.[35] The bottom layer of copper was etched away using ammonia persulphate, the stack gently washed with deionised water, and then the graphene encapsulated $Mo_{1-x}W_xS_2$ monolayer flakes were transferred to a transmission electron microscopy (TEM) support grid. STEM analysis was performed in a JEOL ARM200F TEM with CEOS probe and image aberration correctors at 80 kV. The annular dark-field (ADF) images were recorded at a probe current of ~23 pA and a convergence semi-angle of ~25 mrad using a JEOL annular field detector with an inner and outer collection semi-angle of 45 and 180 mrad, respectively.[36, 37] The scanning rate was typically 20 μs per pixel and each image consists of 2048 × 2048 pixels.[36]

*Determining the atom positions*. Image analysis of the STEM data was used to analyse the atomic configuration in the alloys. Thresholding, based on integrated intensity within a region centred on each atom, was applied to determine the atomic identities and this was then verified visually. The number of W-W pairs at *n*th-nearest-neighbour distances was counted for *n* = 1 to 18, based on the positions of the W atoms in the alloy. Next, the total number of transition metal-transition metal pairs at each distance was counted. The number of W-W pairs is divided by the product of the total number of transition metal-transition metal pairs and the W atom composition *x*.

*Electronic structure measurements.* Angle resolved photoemission spectroscopy (ARPES) data were acquired at the Spectromicroscopy beamline[38] of the Elettra synchrotron, using 27 eV photon illumination and a sample temperature of around 100 K. Spectra were acquired from crystals which were cleaved in UHV immediately before analysis. Three-dimensional data sets of the photoemitted intensity as a function of kinetic energy and emission angles were acquired around the line in reciprocal space along the high symmetry direction from $\bar{\Gamma}$ to $\bar{K}$. From this, 2D energy-momentum slices were extracted showing the dispersion from $\bar{\Gamma}$ to $\bar{K}$. The dispersion of the upper valence band was determined by fitting the corresponding peaks in the energy distribution curves (EDC) with Lorentz functions. The effective mass was found by fitting a parabola to the resultant band dispersions around $\bar{K}$.

*Polarisation-resolved low-temperature photoluminescence spectroscopy.* The PL was excited using a CW dye laser with a wavelength of 625 nm via a 50x 0.55 NA microscope objective producing a 2 μm laser spot. The power incident on the sample was kept below 100 μW. The measurements were performed in backscattering geometry with the sample in vacuum at 4 K. In order to ensure good circular polarisation purity, a quarter wave plate (Thorlabs AQWP05M-600) was placed just before the objective so that in the rest of the optical system the input and output polarisations corresponded to vertical and horizontal linear polarisations. The output polarisation was determined using a linear polariser (Thorlabs LPVIS100) and a half wave plate (AHWP05M-600) used to ensure a constant input polarisation for the spectrometer. The spectrometer used was a Princeton Instruments TriVista 555 with a liquid nitrogen cooled CCD. Measurements were repeated on multiple areas of the sample.

*Monte Carlo simulations.* We model the equilibrium distribution of the cations W and Mo using an Ising-like model of nearest-neighbour interactions between pairs of cations on a hexagonal lattice. The model comprises a single layer with 120×120 in-plane unit cells, a size corresponding to several times the largest apparent feature size in the final distributions. This was initialised to a random distribution constructed to achieve the target concentration *x*. The growth process was simulated by Monte Carlo sampling on proposed site swaps to achieve an equilibrium of the system after 20000 cycles at a temperature of 800 K. To parameterise the Monte Carlo simulations, DFT calculations were performed on a selection of models, as shown in SuppMat Figure S3(a), to determine the optimal value of the interaction energy for a pair of W atoms on adjacent lattice sites.[39] To validate this, DFT calculations were performed for a selection of atomic configurations extracted from those observed in the experimental STEM images (see SuppMat Figure S3(c-d)). For illustration and comparison, Monte Carlo simulations were also performed with interaction energies of -50, 0 and 50 meV.

*Plane-wave DFT.* For calculations of the optimized lattice constants of the pure materials, $MoS_2$ and $WS_2$, the plane-wave DFT package CASTEP was used.[40] A cut-off energy of 424.5 eV and a 10×10×1 k-point grid were used, and the generalized gradient approximation of Perdew, Burke and Ernzerhof (PBE) was employed[41]. Since the resulting lattice constants for $MoS_2$ and $WS_2$ are very similar (within 0.3%), the same value, 3.18 Å, was subsequently used for both materials, thereby neglecting the very small lattice mismatch. CASTEP calculations were also used to provide the interaction energies for pairs of W atoms for the Monte Carlo parameterization, with commensurate adjustments of the k-point grid to 3×3×1 and 1×1×1 as the supercell size increased to 4×4×1 and 12×12×1 respectively.

*Linear-scaling DFT.* The ONETEP linear-scaling density functional theory package[42] was applied to simulate large supercells with random alloy configurations. ONETEP uses a representation of the single-particle density matrix in terms of a minimal number of support functions denoted non-orthogonal generalized Wannier functions (NGWFs). The NGWFs are themselves represented in terms of a grid of psinc functions[43] (a systematic basis equivalent to plane waves), which here use a cut-off energy of 1200 eV, and an NGWF cut-off radius of $10a_0$. The PBE functional was used, with projector-augmented wave (PAW)[44,45] potentials from the JTH library[46]. The electronic properties of different random configurations of atomic structures were calculated within an 8×8×1 supercell model. $Mo_{1-x}W_xS_2$ monolayer models were created for various concentrations, $x$, with cation distributions determined according to the procedures described below. A separate geometry optimisation was performed for each structure, with force tolerance of 0.1 eV/Å. Since the band structures obtained from the disordered alloy supercell calculations have a heavily-folded band structure that is not directly comparable to the ARPES results, the effective band structure was calculated in the primitive cell using the spectral function unfolding method[14, 47], the adaptation of which to the NGWF representation is described in Ref [48].

*Virtual crystal approximation (VCA) calculations.* Calculations were performed within the Virtual Crystal Approximation[49] with the Quantum Espresso code.[50,51] PAW datasets representing the W and Mo potentials mixed in the appropriate ratio $x$ were constructed using the 'virtual' tool within the Quantum Espresso package. Spin-orbit coupling was included using the treatment described by dal Corso.[52–54]

## 3. Results and Discussion

$Mo_{1-x}W_xS_2$ alloy crystals were grown by chemical vapour transport, as described in Methods, across the range from $x = 0$ to 1. Their compositions, determined by a combination of XPS and EDX analysis (see Methods and SuppMat S2 for further details), were found to be consistent within 3% between crystals grown in the same batch, and across individual crystals when averaged at the micrometre scale. When exfoliated to monolayers, they exhibit room temperature photoluminescence with peak emission frequency varying nonlinearly from 1.85 eV at $x = 0$ to 1.98 eV at $x = 1$, consistent with prior literature reports [14] (SuppMat S4).

### 3.1 Atomic structure

To determine whether the alloys are atomically ordered, atomic resolution ADF-STEM images of monolayer $Mo_{1-x}W_xS_2$ flakes were acquired. Example images are given in Figure 2a. In the magnified view, Figure 2b, the transition metal atoms can be resolved but the chalcogen atoms give only weak contrast, confirmed by the corresponding simulated STEM image of the same area in Figure 2c. Due to the large difference between their atomic numbers, there is also a clear contrast difference between Mo (Z=42) and W (Z=74), again verified by comparison between the line profiles of the experimental and simulated images, as shown in Figure 2d.

For comparison, we performed Monte Carlo simulations made with varying interaction energies, $J$, and compositions, $x$. Details of the simulation methodology are given in the methods and SuppMat S3. To predict $J$ for these alloys, a range of atomic configurations were tested, from which the binding energy of substitutional W atoms within the $MoS_2$ monolayer was found to be well-represented by an average value of $J = 7.6$ meV. This was further verified by comparison of the Monte Carlo model to DFT calculations of a range of larger cells (SuppMat S3). Figure 3a shows representative atomic arrangements from Monte Carlo simulations, comparing atomic arrangements for the predicted value of $J = 7.6$ meV to those with stronger interactions ($J = \pm50$ meV) at a simulation temperature of 800 K. At negative $J$, clusters of W tend to form, while at positive $J$, isolated W atoms are energetically preferred.

For $J=0$, when there is no energy difference between the W and Mo, they are randomly arranged; qualitatively, Monte Carlo simulations with $J = 7.6$ meV look similar to those with $J = 0$. Although close visual inspection of the images can suggest some ordering, quantitative statistical analysis is required to rigorously test whether the Mo/W atoms are randomly distributed within the sheets. SuppMat S3 shows that in the Monte Carlo model, strong ordering disappears beyond 50 K and is completely absent for temperatures comparable to the growth conditions. For a given composition, ADF-STEM images with areas of 120×120 atoms were analysed, fitting the intensity at each atom coordinate and thresholding to determine the arrangement of Mo and W atoms, as described in Methods. With the atom coordinates identified, statistical analysis can be used to investigate whether the arrangements are ordered. Following Cowley [55], we define the short range order parameter: $\alpha_i = 1 - n_i/(x\,c_i)$, where $n_i$ is the number of W atoms out of the $c_i$ atoms in the $i$th coordination shell surrounding each W atom (as defined schematically in Figure 1), and $x$ is the proportion of W atoms (to $(1 - x)$ Mo atoms) in the alloy.

Figure 3e shows the order parameters for $Mo_{0.78}W_{0.22}S_2$ determined from the ADF-STEM images alongside those calculated from analysis of the Monte Carlo simulations. Order parameters for the other compositions followed similar trends. For the simulated data, if $J > 0$ then $\alpha_1 < 0$, showing a lower probability of finding a W atom neighbouring another W atom. The magnitude of $\alpha_i$ decreases rapidly with increasing $i$, indicating a finite range to the order. If $J < 0$ then $\alpha_1 > 0$ (a tendency to cluster), and $\alpha_i$ decreases monotonically to 0 with increasing $i$. The order parameters determined from the experimental data agree, within uncertainties, with data from both $J = 0$ and $J = 7.6$ meV simulations, consistent with a random distribution of W atoms. We conclude that under the conditions used for the growth of these crystals, the resultant atomic distributions are indistinguishable from a random distribution and are consistent with those expected at thermodynamic equilibrium at the elevated growth temperature of this system.

### 3.2 Electronic structure

To correlate the electronic structure to the atomic structure, the valence band dispersions of the $Mo_{1-x}W_xS_2$ alloy crystals were measured by ARPES. Since the measurements are surface sensitive, the as-grown bulk crystals were mounted flat onto a sample plate and cleaved in the ultrahigh vacuum (UHV) chamber immediately prior to analysis, to generate the clean surface required. Cleaving naturally results in exposure of the (002) planes to the beam, so that ARPES probes the electronic structure in the plane of the 2D layer.

Typical spectra are given in Figure 4, showing the dispersion from $\overline{\Gamma}$ to $\overline{K}$ for three alloy compositions: $Mo_{0.63}W_{0.37}S_2$, $Mo_{0.5}W_{0.5}S_2$ and $Mo_{0.29}W_{0.71}S_2$. We concentrate our analysis on the structure around the zone corners at $K$ where, for the monolayers, the VBM is found. Around $K$, the spin-split upper valence band has in-plane orbital composition from the metal $d_{x2-y2} + d_{xy}$ and, to a lesser extent, the chalcogen $p_{x+y}$.[56] This results in a weak $k_z$ dispersion such that the key features of the spectra in that region are quantitatively similar for bulk and monolayer materials.[57, 58] From this dispersion, band parameters can be determined such as the spin-orbit splitting of the bands at $K$, the effective mass, and the band width (see the schematic in Figure 5).

The qualitative trends are apparent by visual inspection of the ARPES spectra: as the W content increases, the spin-orbit splitting between the upper valence bands near $K$ ($SOC_{vb}$) increases; the band width ($D_{vb}$) increases and the effective mass decreases. To quantitatively evaluate the evolution of the band structure with composition, measurements were performed on all the compositions (x = 0.22, 0.37, 0.49, 0.71, 0.87 and 1) and band parameters extracted quantitatively by fitting the ARPES spectra, as described in Methods. The data are presented in Figure 6, and compared to first principles calculations as described below.

*Ab initio* calculations of the electronic structure across the alloy compositions were made using two approaches: the virtual crystal approximation (VCA),[49] where fractional atom compositions are considered within the primitive cell of pure materials; and the more computationally intensive approach of calculating the electronic structure of large supercells with varying ratios of Mo:W atoms. The details of each approach are described in Methods and Supplementary Materials. In brief, the Quantum Espresso package[50,51] was used for the VCA calculations, and linear-scaling density functional theory (LS-DFT) calculations were performed on $8 \times 8 \times 1$ supercells (192 atoms) by the ONETEP package.[59] The LS-DFT calculations were unfolded into the primitive cell of the pure monolayers[48]; an example spectral function, for $Mo_{0.5}W_{0.5}S_2$, is shown in Figure 5, and spectral functions for the other concentrations can be seen in SuppMat Figure S7. Band structure parameters were determined from fitting the spectral functions.

Figure 6 compares the band structure parameters measured from the experimental data to those calculated by LS-DFT and VCA simulations, as a function of the W content in the alloy. As the ARPES spectra probe the valence band, direct experimental measurements of the conduction band edge and band gap ($E_g$) are not possible from this data. Instead, the predicted $E_g$ are compared to the PL emission peak energy in Figure 6a. DFT calculations normally underestimate $E_g$,[6] so we concentrate on the trends with composition. As previously reported,[14] the PL peak energy changes non-linearly with composition, with the bowing parameter here determined to be $0.17 \pm 0.01$ eV. This non-linear dependence of $E_g$ on $x$ is expected due to the change in orbital characteristic of the states at the conduction band edge from $MoS_2$ to $WS_2$.[14] LS-DFT predicts qualitatively similar changes in $E_g$, with a bowing parameter of $0.12 \pm 0.01$ eV, but with a smaller change in $E_g$ ($E_g = 1.63$ eV for $MoS_2$ and 1.62 eV for $WS_2$) than observed in the change in PL peak emission energy (1.85 eV for $MoS_2$ increasing to 1.98 eV for $WS_2$). The magnitude of $E_g$ determined from VCA is similar to the LS-DFT calculations, but the bowing parameter becomes negative (-$0.02 \pm 0.01$ eV), showing that the VCA predictions deviate from the experimental observations.

Caution should be taken when comparing the PL emission peak energies to the predicted band gaps ($E_g$) which are determined from the energy difference between the VBM and the CBM. The large exciton binding energy (several hundred meV) means that the PL peak emission energy is not a good measurement of the single-particle electronic band gap.[2] Interpretation of the PL is further complicated by the change in order of the spin-polarised conduction and valence bands, as shown schematically in Figure 5b. The PL emission is from the A exciton (bright exciton) in the $K$ valley, corresponding to an electron transition between states of the same spin. For $x = 0$ ($MoS_2$), this corresponds to the VBM and CBM states. But at $x = 1$ ($WS_2$), this has changed and the bright A exciton emission comes from the second-lowest lying conduction band which is greater in energy than the CBM by the spin-orbit splitting at the conduction band edge (around 30 meV for $WS_2$).

Prior direct measurements of $E_g$ from ARPES of electrostatically doped monolayers found $E_g = 2.07 \pm 0.05$ eV for $MoS_2$ and $E_g = 2.03 \pm 0.05$ eV for $WS_2$,[60] only a small change in band gap as also seen in the LS-DFT results. The difference between the LS-DFT and PL data then suggests a consistent decrease in exciton binding energy as the W content increases.

ARPES provides a direct and visual measurement of the valence band dispersion of the alloys, allowing easier comparison to calculated spectra than is possible from optical spectroscopy data. Measured values of $SOC_{vb}$ from the $Mo_{1-x}W_xS_2$ single crystals (blue data points) are compared in Figure 6b to those calculated for the monolayer alloys by LS-DFT (black) and VCA (red). The measured value for $WS_2$ ($SOC_{vb} = 458 \pm 10$ meV) is consistent with prior reports for the bulk crystals ($450 \pm 30$ meV [39]), and for a monolayer flake of $WS_2$ ($450 \pm 30$ meV [60]). For the experimental results, $SOC_{vb}$ increases linearly as a function of the W content, within the error bars. The LS-DFT predictions show a similar linear trend, with a systematic under-estimate of the experimental results by around 50 meV. The VCA calculations of $SOC_{VB}$ are similar in magnitude to the experimental measurements but show a non-linear dependence on composition. A LS-DFT simulation was also made using the atom positions determined directly from a STEM image: the band parameters for this simulation are shown as the purple data points in the graphs in Figure 6 and are consistent with the other LS-DFT simulations. The band width ($D_{vb}$), Figure 6c, follows similar trends to $SOC_{VB}$: the agreement between the experimental data and the LS-DFT data is excellent, other than a systematic offset (under-estimate) between the two; and the VCA data shows a bowing that is not present in either the LS-DFT or experimental data.

Figure 6d compares the evolution of the valence band effective mass (around **K** and in the **Γ** to **K** direction) calculated from ARPES spectra, the LS-DFT and VCA calculations. It is difficult to accurately determine the effective mass from the ARPES spectra, as a result the uncertainties are large and there are no clear trends. From the DFT calculations it is clear that the effective mass decreases monotonically with $x$, consistent with the increasing curvature of the bands and increasing band widths. This decrease in effective mass is also consistent with a decrease in exciton binding energy with increasing W content.

The ARPES spectra do not give information about the conduction band as they only probe the occupied states. However, the spin-orbit splitting at the conduction band edge ($SOC_{CB}$) can be extracted from the DFT spectral functions and this is plotted as a function of composition in Figure 6e. It is an order of magnitude lower than $SOC_{VB}$. At small $x$, closer to pure $MoS_2$, the order of the conduction bands changes, see Figure 5b. Hence $SOC_{CB}$ initially decreases to zero and then increases with $x$. Although the LS-DFT and VCA calculations do not quantitatively agree over the composition at which $SOC_{CB} = 0$, it is clear that at intermediate compositions $SOC_{CB}$ is small. At $x=0$ and $x=1$, the small differences between the results are due to the different approaches to SOC taken by the two simulation packages: ONETEP uses a perturbative approach to SOC, whereas Espresso uses a full SOC treatment integrated into the self-consistency procedure.

Calculations were performed on three different random realisations of the 8×8 supercells for each $x$, allowing a rough estimate of the disorder potential to be made, on a similar lengthscale to the exciton radius, based on the variation of the CBM position between these random realisations. From these we observe that the disorder potential is larger than $SOC_{CB}$ at least for $0.125 \leq x \leq 0.5$, becoming smaller again as $SOC_{CB}$ increases (see SuppMat S5). This casts doubt over the usual picture of separated states split by SOC at the conduction band edge and merits further investigation of the spin-valley locking in these alloys.[61]

### 3.3 Spin-valley locking

To test whether spin-valley locking is retained in the alloys, optical spectroscopy measurements were performed on an exfoliated heterostructure consisting of a monolayer flake of $Mo_{0.5}W_{0.5}S_2$ encapsulated in hBN. An optical image, and schematic, of the sample is shown in Figure 7a. Polarisation-resolved PL spectra, Figure 7b, show clear circular dichroism: excitation with righthand circularly polarised light ($\sigma^+$) leads to emission of mainly $\sigma^+$ light, whilst $\sigma^-$ excitation leads to $\sigma^-$ emission. Excitation with vertical linear polarisation (corresponding to $(\sigma^+ + \sigma^-)/\sqrt{2}$) results in emission with an equal amount of $\sigma^+$ and $\sigma^-$ circular components. The PL spectra are plotted with normalised intensity to allow direct comparison of the excitations with different polarisations. The valley polarisation, $\eta = \frac{PL(\sigma^+) - PL(\sigma^-)}{PL(\sigma^+) - PL(\sigma^-)}$ where $PL(\sigma\pm)$ is the intensity of the corresponding circularly polarised component of the PL emission,[9] is found to be $\eta = 0.47$ under $\sigma^+$ excitation and under $\sigma^-$ excitation. These results are fully consistent with the expectations from the optical selection rules for the spin-polarised monolayer materials and with previous observations of spin-valley locking in pure monolayer MoWSeS flakes,[5-8] showing that spin-valley locking is retained for the monolayer alloys despite the atomic-scale heterogeneity in the alloys and the small $SOC_{CB}$.

## 4. Conclusions

No ordering or segregation was found in CVT grown $Mo_{1-x}W_xS_2$ alloys, consistent with Monte Carlo simulations parameterised by ab initio calculations. ARPES measurements showed that the spin orbit coupling at the valence band edge increases linearly with increasing $x$ (increasing W content). LS-DFT simulations using large unit cells gave predictions consistent with the ARPES measurements, but DFT simulations using fractional atom concentrations in a primitive unit cell were found to be inconsistent with the experimental results. This conforms to expectations of the known severe limitations of the predictive power of VCA, which becomes even more apparent for properties that are nonlinear with atomic number Z, such as SOC. This is especially apparent for materials which retain some localised character in the valence band, and strongly suggests instead the need for large-scale disordered models to study alloy properties. From the LS-DFT models, for intermediate alloy compositions the spin orbit coupling is predicted to be smaller than the spatially localised changes in the conduction band edge due to the atomic-scale heterogeneity of the alloy. Despite this, polarisation-resolved PL measurements of a monolayer $Mo_{0.5}W_{0.5}S_2$ flake show circular dichroism consistent with spin-valley locking. The ability to rationally and continuously tune the spin-orbit coupling, effective mass and band gap suggests a promising future for MX$_2$ alloys in valleytronic and spintronic applications.

## Acknowledgements

David Quigley is thanked for the provision of a Monte Carlo model code which was adapted for the current work. The Engineering and Physical Sciences Research Council supported studentships for NCT (EP/M508184/1) and AJG (EP/R513374/1), and funded NW and NDMH through EP/P01139X/1. XXia and SML were supported by University of Warwick Chancellor's Scholarships. Computing resources were provided by the High Performance Computing Service, the Scientific Computing Research Technology Platform of the University of Warwick, and the UK national high performance computing service, ARCHER, via the UKCP consortium (EP/P022561/1). The research leading to this result has been supported by the project CALIPSOplus under Grant Agreement 730872 from the EU Framework Programme for Research and Innovation HORIZON 2020.

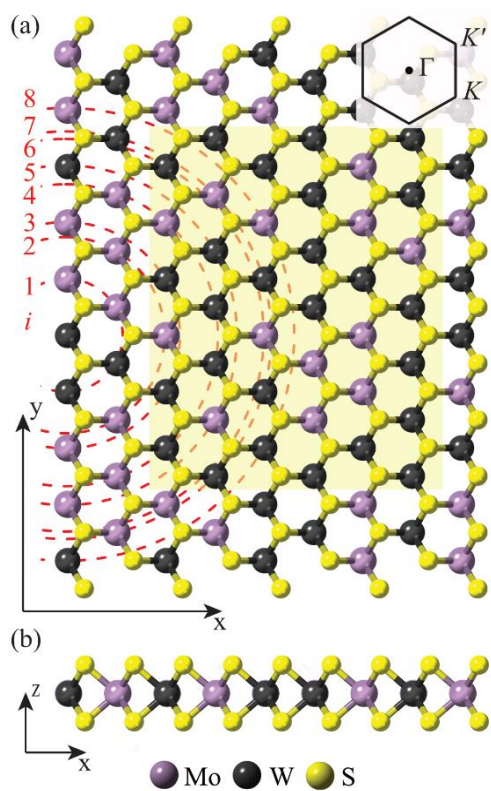

Figure 1. Atomic schematic of the structure of monolayer $Mo_{1-x}W_xS_2$. (a) top-view with co-ordination shells overlaid for $i = 1$ to 8 (red dashed lines); inset at the top right is a schematic of the first Brillouin zone. (b) side vew.

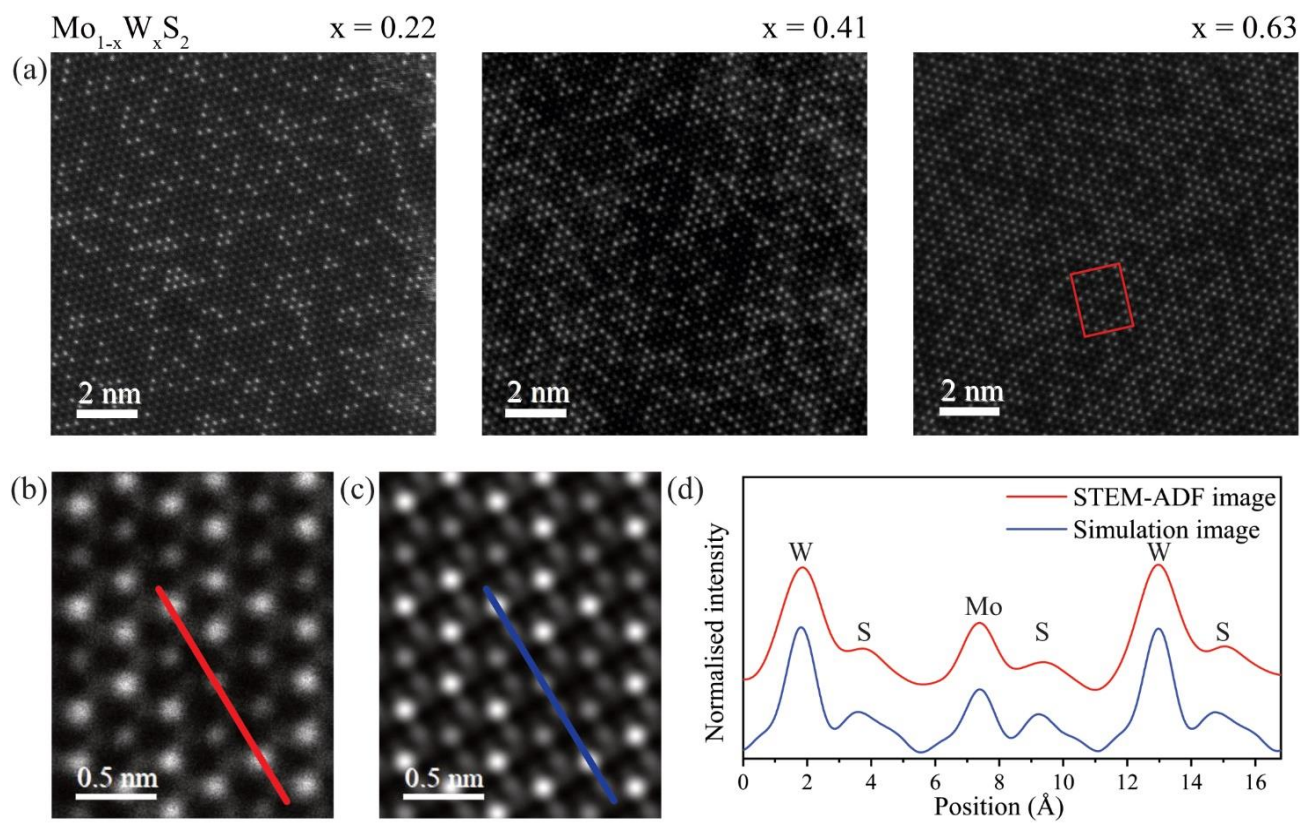

Figure 2. Atomic structure of monolayer $Mo_{1-x}W_xS_2$ revealed by STEM-ADF. (a) STEM-ADF images of monolayers with *x*=0.22, 0.41 and 0.63 as labelled. (b) Magnified image of the area highlighted by the red rectangle in (a). (c) Simulated ADF image for the region in (b). (d) Intensity profiles along the red and blue line in experimental and simulated ADF images, respectively.

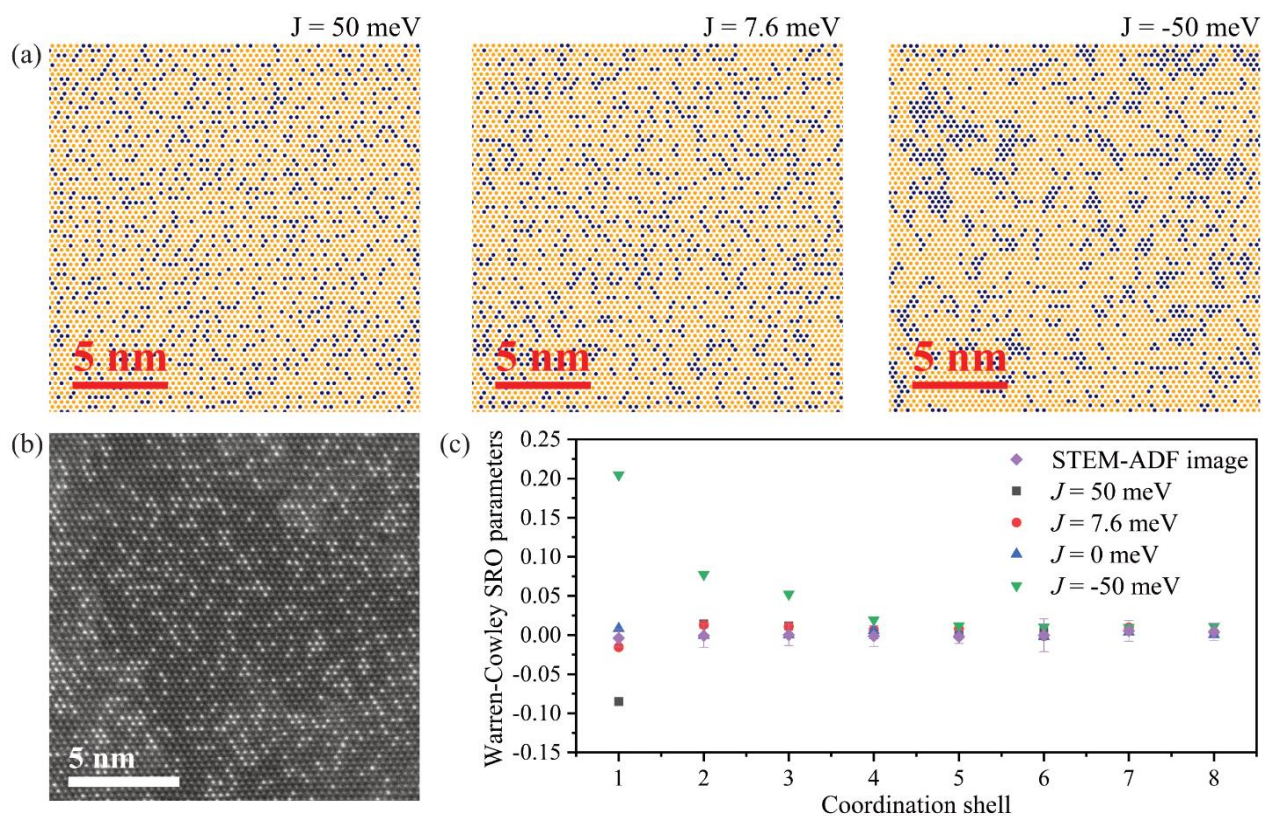

Figure 3. Statistical analysis of the atomic arrangements in $Mo_{1-x}W_xS_2$. (a) Monte Carlo simulations for $x$ = 0.22 with binding energy $J$ = 50, 7.6 and -50 meV, respectively. (b) STEM-ADF image of monolayer $Mo_{0.78}W_{0.22}S_2$. (e) Comparison of the Warren-Cowley SRO parameters calculated from the Monte Carlo simulations to those determined from the experimental ADF-STEM images, for $x$ = 0.22.

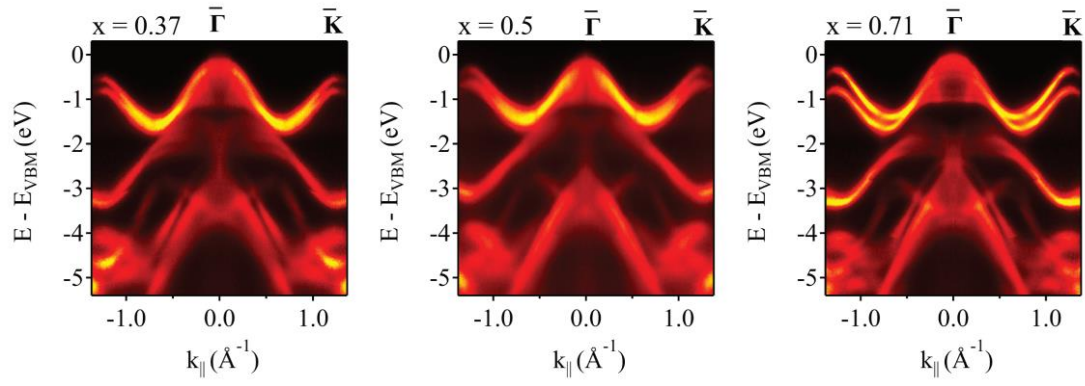

Figure 4. ARPES energy momentum slices along the $\bar{\mathit{\Gamma}}$ to $\bar{\mathit{K}}$ direction for crystals of (from left to right) $Mo_{0.63}W_{0.37}S_2$, $Mo_{0.5}W_{0.5}S_2$ and $Mo_{0.29}W_{0.71}S_2$. The spectra are mirrored around $\bar{\mathit{\Gamma}}$ for clarity.

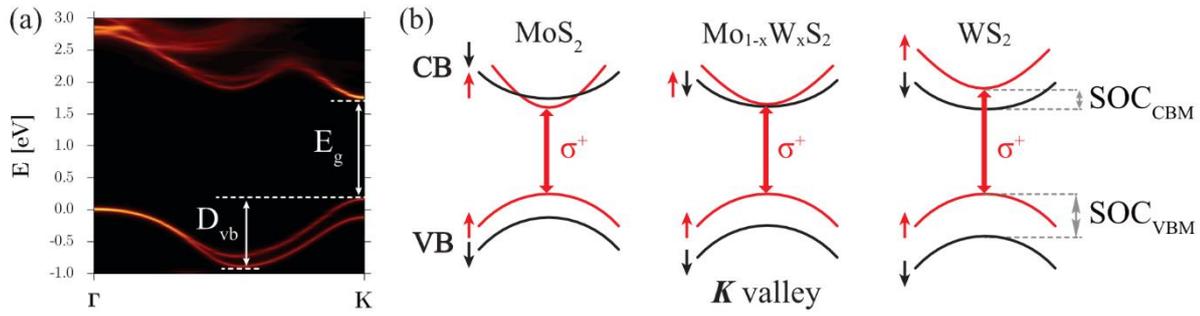

Figure 5. Band parameters from the band dispersions. (a) LS-DFT spectral function calculation for monolayer $Mo_{0.5}W_{0.5}S_2$ with band gap ($E_g$) and bandwidth ($D_{vb}$) definitions overlaid. (b) Schematics of the spin-polarised band edges at **K** for monolayer $MoS_2$, $Mo_{1-x}W_xS_2$ and $WS_2$, showing that the order of the conduction band reverses from $MoS_2$ to $WS_2$ and is almost degenerate at intermediate compositions.

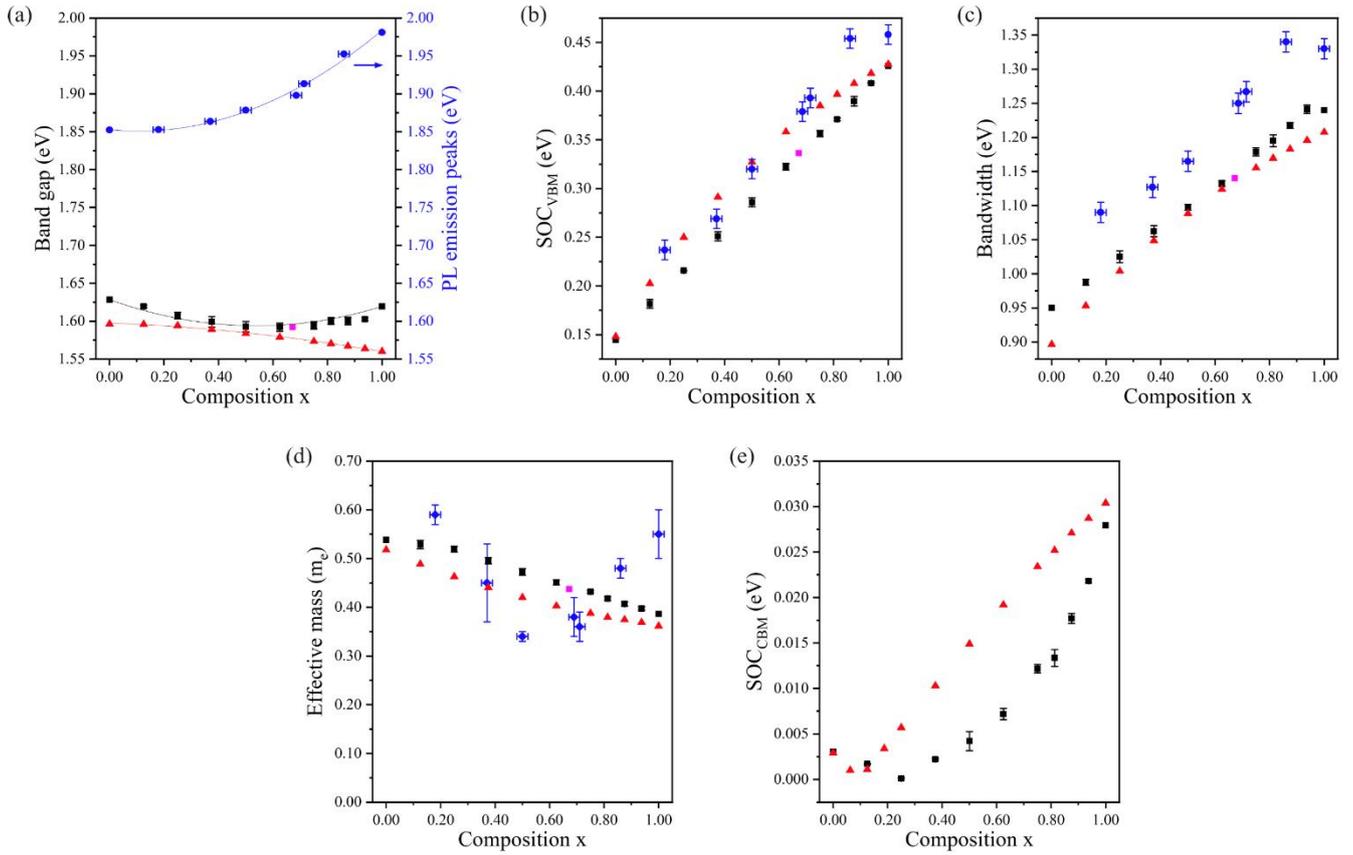

Figure 6. Comparison of band structure parameters determined from experimental data and DFT calculations, as a function of the composition $x$ in $Mo_{1-x}W_xS_2$: (a) the evolution of PL emission peak energy and DFT calculated band gaps with $x$; (b) SOC at the valence band edge, $SOC_{VBM}$; (c) band width, $D_{vb}$; (d) effective mass of the upper valence band at K, in the $\Gamma$ to K direction; and (e) SOC at the conduction band edge, $SOC_{CBM}$. The data are from experimental measurements (blue data points), VCA calculations (red data points) and LS-DFT calculations (black data points). The purple data points are from LS-DFT calculations using atomic arrangements determined from a STEM-ADF image.

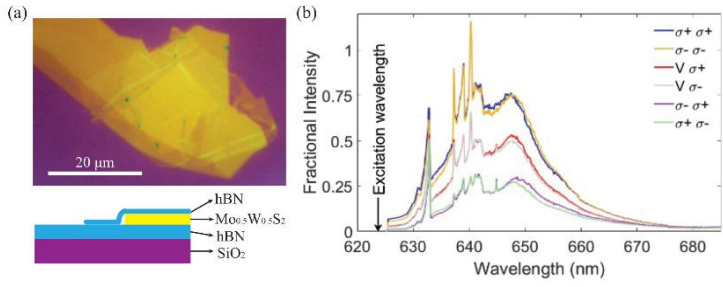

Figure 6. Polarisation-resolved PL spectroscopy of monolayer $W_{0.5}Mo_{0.5}S_2$. (a) Optical microscopy image of the hBN encapsulated $W_{0.5}Mo_{0.5}S_2$ mono-flake; bottom, schematic cross-section of the sample. (b) Polarization resolved photoluminescence spectra of the encapsulated sample taken at 4 K. $\sigma^+$ and $\sigma^-$ correspond to light with spin parallel and antiparallel to the direction of the incoming beam. V corresponds to vertical linear polarization *i.e.* has spin $((\sigma^+ + \sigma^-))/\sqrt{2}$. The pairs of spectra with the same incident polarization have been scaled by the sum of their intensities at 646 nm to provide an indication of the fraction of the intensity in each polarization. No significant variation in the sum of the intensities was observed.

Supplementary Material for:

# Atomic and electronic structure of two-dimensional Mo$_{(1-x)}$W$_x$S$_2$ alloys

**Xue Xia[1§], Siow Mean Loh[1§], Jacob Viner[2], Natalie C. Teutsch[1], Abigail J. Graham[1], Viktor Kandyba[3], Alexei Barinov[3], Ana M. Sanchez[1], David C. Smith[2], Nicholas D. M. Hine[1*], Neil R. Wilson[1*]**

[1] Department of Physics, University of Warwick, Coventry, CV4 7AL, UK
[2] School of Physics and Astronomy, University of Southampton, Southampton SO17 1BJ, UK
[3] Elettra - Sincrotrone Trieste, S.C.p.A., Basovizza (TS), 34149, Italy

§ these authors contributed equally

E-mail: neil.wilson@warwick.ac.uk, n.d.m.hine@warwick.ac.uk

## Table of Contents

1.  Crystal growth of $Mo_{1-x}W_xS_2$ by chemical vapour transport.

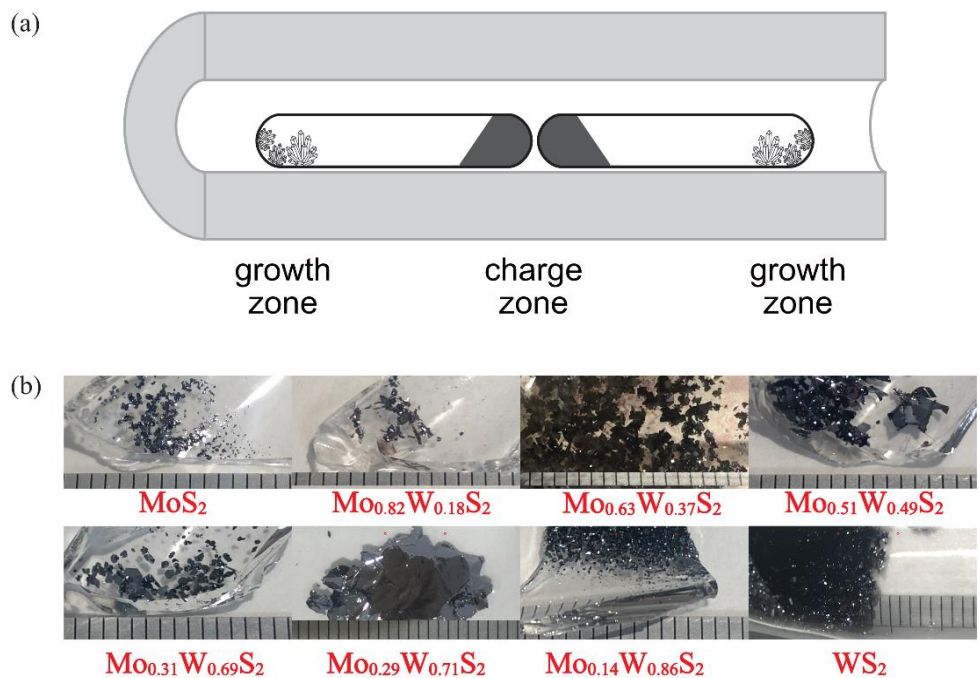

Figure S1. (a) Schematic of the chemical vapor transport setup. (b) Photographs of as-grown $Mo_{1-x}W_xS_2$ bulk crystals, $x$ = 0, 0.18, 0.37, 0.49, 0.69, 0.71, 0.86 and 1. The ruler at the bottom of each photograph shows millimetre graduations.

Single crystals were synthesised by chemical vapor transport (CVT) in a two-step process. Mo, W and S element powders (all purity 99.9%, Sigma-Aldrich) were mixed stoichiometrically into an ampoule which was pumped down to a pressure of $10^{-6}$ mbar and sealed. The ampoule was heated to 1000 °C for 3 days to form $Mo_{1-x}W_xS_2$ powder. Some of this powder (about 2 g) was transferred to a new quartz ampoule with larger inner diameter ( 1.6 cm) and mixed with the transport agent, $I_2$ (10 mg cm$^{-3}$ of the ampoule volume). To keep the $I_2$ stable, the ampoule was evacuated to $10^{-6}$ mbar and placed in ice while sealing. The ampoule was then placed into a three-zone furnace as shown in Figure S1a; the charge zone was kept at 1050 °C for 20 days with the growth zone at 950 °C. After cooling to room temperature, the ampoule was opened in air and single crystals were collected at the growth end, Figure S1b.

2. Compositional analysis

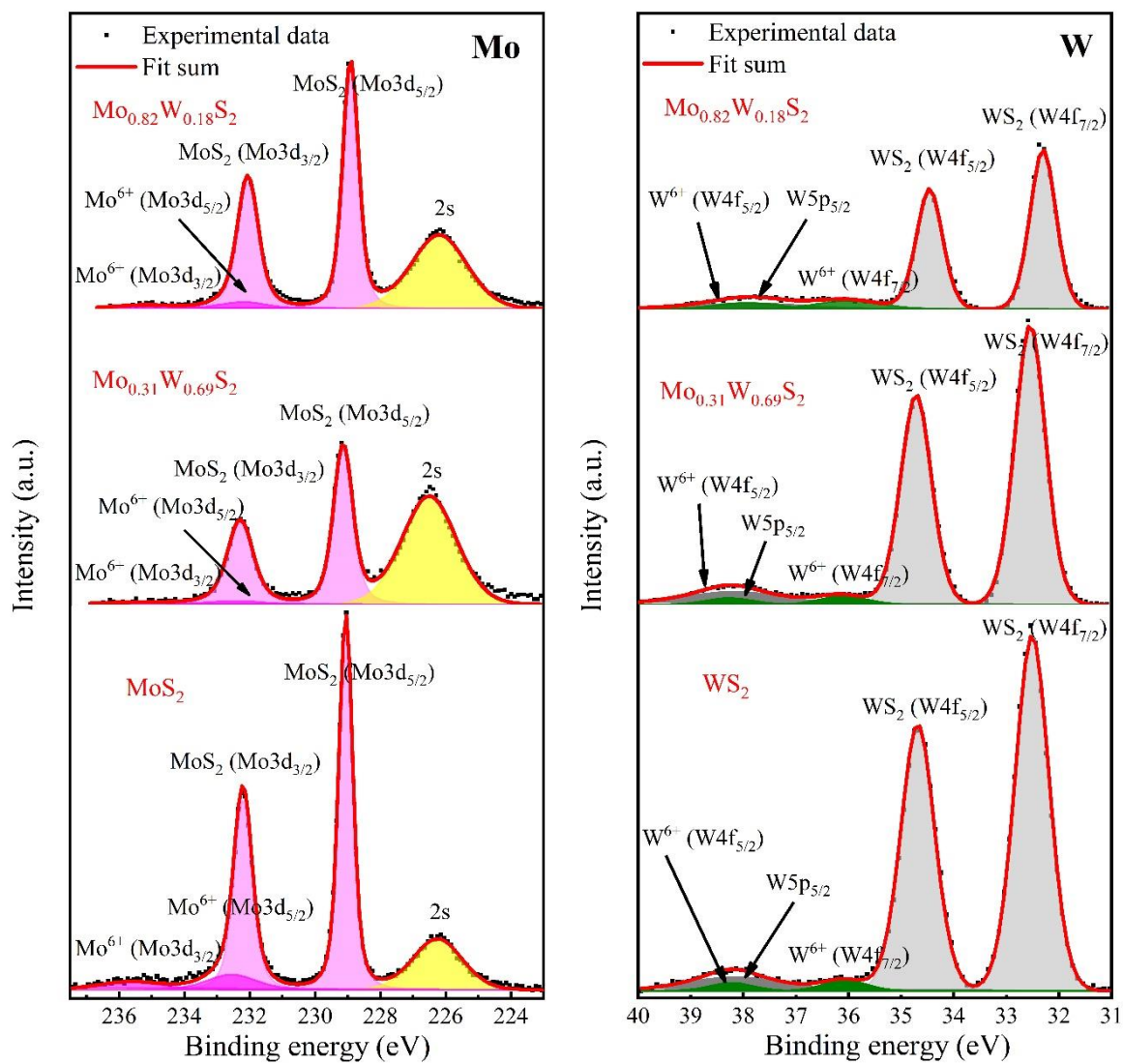

Figure S2. X-ray photoelectron measurements on $Mo_{1-x}W_xS_2$ single crystals ($x$ = 0, 0.18, 0.69 and 1). Mo 3d and W 4f spectra with Gaussian-Lorentzian fitting of each peak.

Figure S2 shows example XPS spectra from alloy $Mo_{1-x}W_xS_2$ single crystals. For pure 2H-$MoS_2$, the Mo $3d_{3/2}$ and Mo $3d_{5/2}$ core levels are located at 232.0 eV and 228.9 eV, respectively, while for 2H-$WS_2$ crystals the W $4f_{5/2}$ and W $4f_{7/2}$ peaks are at 34.5 eV and 32.3 eV, respectively. There are additional doublet features at higher binding energies in the crystals that are associated to their oxidation states (Mo-O and W-O), respectively. Compared to the pure materials, in the pristine 2H-$Mo_{0.31}W_{0.69}S_2$, the Mo $3d_{5/2,3/2}$ peaks and oxidised states shift to lower binding energies, while shifts of the W $4f_{7/2,5/2}$ peaks of these states are very small. For the Mo rich end, both the Mo $3d_{5/2,3/2}$ and W $4f_{7/2,5/2}$ peaks shift to lower binding energy, compared to the W rich and pure W materials. These observations are consistent with previous reports.[1,2] The stoichiometric ratio of W to (Mo + W) was calculated by integrating the intensities of the Mo and W fitting peaks, applying standard sensitivity-factor corrections. As XPS is surface sensitive, the crystals were also analysed by EDX to check for consistency in composition. SEM-EDX was carried out in a Zeiss Gemini SEM 500 with an accelerating voltage of 5 kV and an Oxford 150 EDS detector. The composition for the different alloys, obtained by SEM-EDX and XPS, are summarised in Table S1.

Table S1. Crystal compositions determined by EDX and XPS for $Mo_{1-x}W_xS_2$ single crystals.

| $Mo_{1-x}W_xS_2$ | Nominal W content | EDX | Error bar | XPS | Error bar |
|---|---|---|---|---|---|
| $MoS_2$ | 0 | 0 | 0.02 | 0 | 0.02 |
| $Mo_{0.82}W_{0.18}S_2$ | 0.13 | 0.18 | 0.02 | 0.16 | 0.02 |
| $Mo_{0.63}W_{0.37}S_2$ | 0.38 | 0.37 | 0.02 | 0.37 | 0.02 |
| $Mo_{0.51}W_{0.49}S_2$ | 0.50 | 0.50 | 0.02 | 0.49 | 0.02 |
| $Mo_{0.31}W_{0.69}S_2$ | 0.63 | 0.69 | 0.02 | 0.69 | 0.02 |
| $Mo_{0.28}W_{0.72}S_2$ | 0.75 | 0.71 | 0.02 | 0.74 | 0.02 |
| $Mo_{0.14}W_{0.86}S_2$ | 0.86 | 0.86 | 0.02 | 0.86 | 0.02 |
| $WS_2$ | 1.00 | 1.00 | 0.02 | 1.00 | 0.02 |

## 3. Monte Carlo simulations

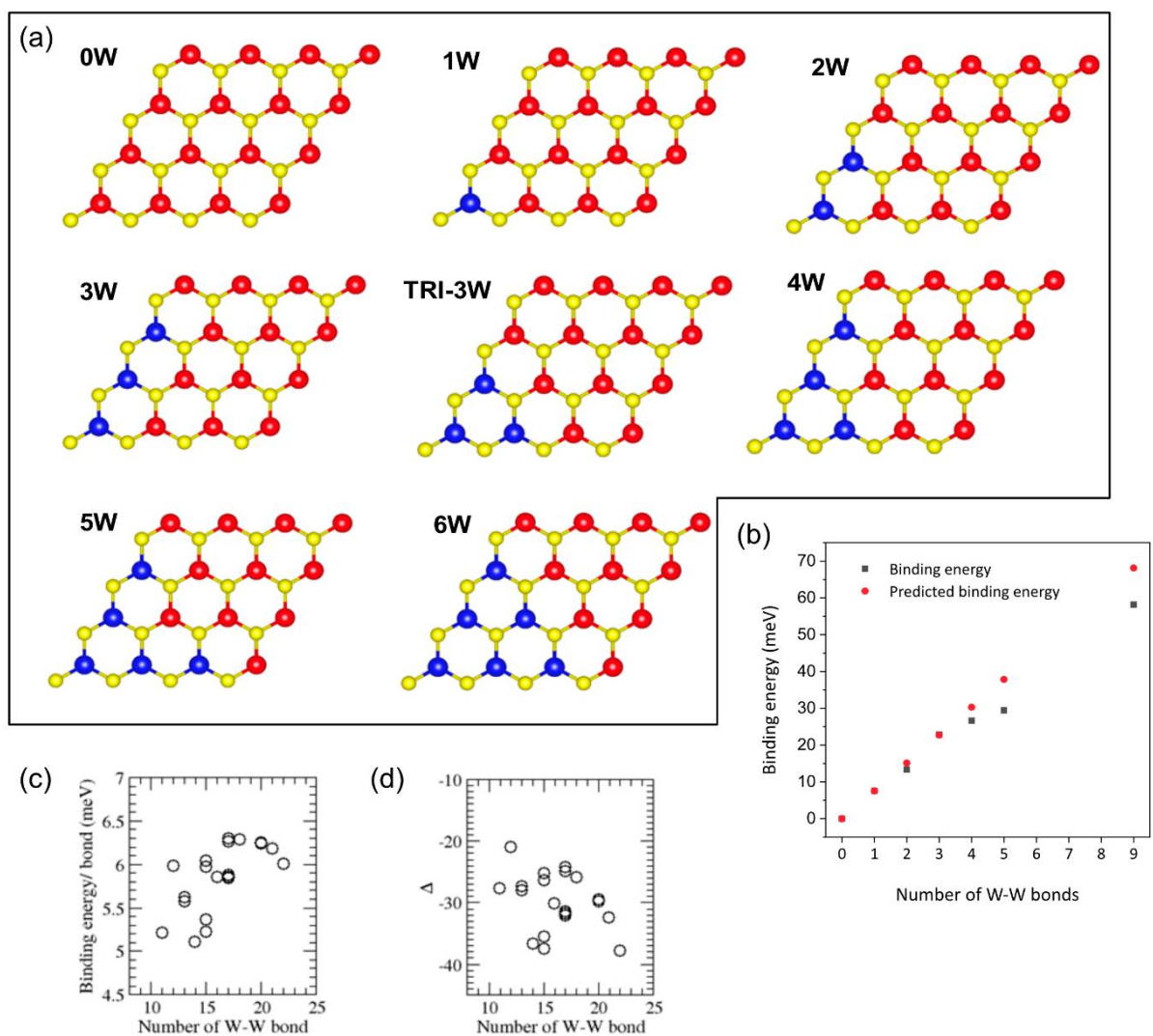

Figure S3. (a) Atomic schematics of the different $Mo_{(1-x)}W_xS_2$ atomic configurations used in the DFT calculations. (b) Comparison between the binding energy calculated by DFT and that predicted using a pairwise interaction calculated from an average value. For twenty $Mo_{0.8}W_{0.2}S_2$ models generated with random configurations within 12×12×1 supercells, we also show (c) the DFT-calculated binding energy and (d) the difference Δ (meV) compared to the value expected based on the number of W-W bond (as would be measured in our Monte Carlo model). These combine to verify that the energy model used in the Monte Carlo calculations represents a reasonable approximation to the DFT results.

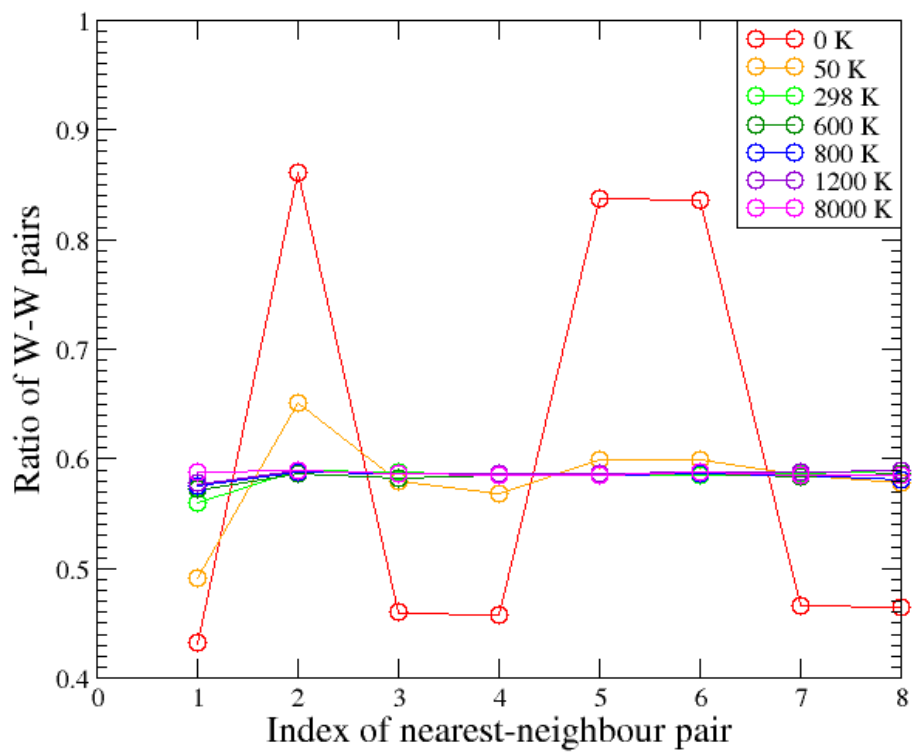

Figure S4. Monte Carlo simulation results showing the ratio of W-W pairs as a fraction of the total number of TM-TM pairs for each nth-nearest neighbour shell, with pairwise interactions between nearest neighbours with J= 7.6 meV. Results are shown for a range of temperatures from T = 0 K to

T = 8000 K. At sufficiently low temperatures (T ≤ 50K), clear evidence is seen of long-ranged ordering, but this ordering has almost vanished by the temperatures that will be encountered during the growth process (T ≈ 800 K).

## 4. <u>Raman spectroscopy and photoluminescence of the $Mo_{1-x}W_xS_2$ alloys</u>

Figure S5. (a) Raman spectra from 2H- $Mo_{1-x}W_xS_2$ single crystals (at 532 nm excitation). (b) The

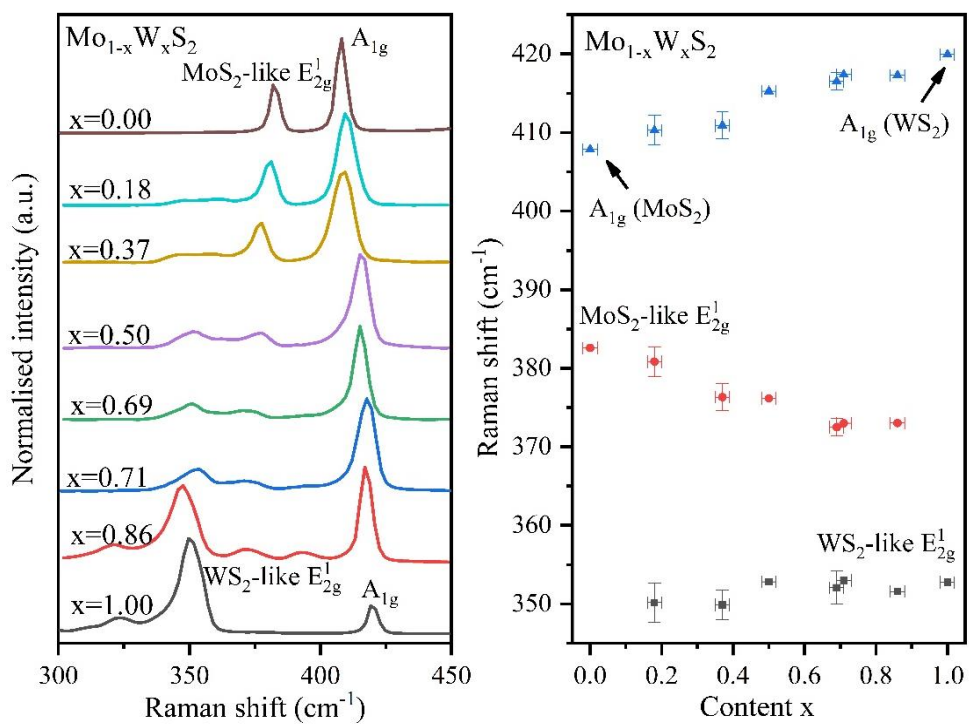

variation of $A_{1g}$ and $E_{2g}^1$ mode peak positions with composition.

Figure S5 shows the changes in Raman spectra in 2H- $Mo_{1-x}W_xS_2$ single crystals as *x* increases. Bulk $MX_2$ (M = Mo and W) possess $D_{6h}$ point group symmetry; they have 12 modes of lattice vibrations at the $\Gamma$ point in the Brillouin zone and four of them are first-order Raman-active, $A_{1g}$, $E_{1g}$, $E_{2g}^1$ and $E_{2g}^2$. In back-scattering experiments on the surface perpendicular to the *c*-axis, $A_{1g}$, $E_{2g}^1$ and $E_{2g}^2$ are allowed.[3] The Raman spectra confirm the high quality of the alloy crystals, and the change in Raman shift with composition is consistent with previous reports.[4]

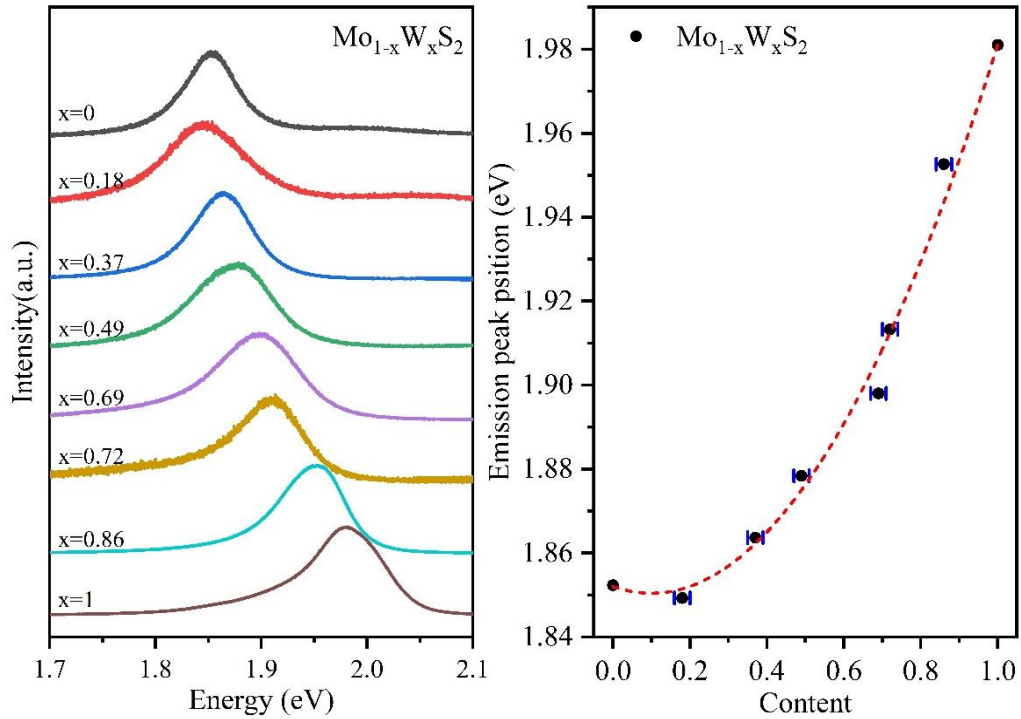

Figure S6. PL emission spectra from $Mo_{1-x}W_xS_2$ monolayers. (a) PL spectra of $Mo_{1-x}W_xS_2$ monolayers with 532 nm excitation. (b) Evolution of the bright A exciton PL emission peak position with increasing W composition *x* (0 to 1), the dashed red curve is a fit to the data, as described below.

Figure S6a shows PL spectra for monolayer flakes of all the compositions from $MoS_2$ to $WS_2$. The dependence of the photoemission peak energy, from the bright A exciton, on composition is shown in Figure S6. The A exciton emission peak shifts continuously and nonlinearly from 1.855 eV at *x* = 0 to 1.981 eV at *x* = 1. The PL peak energy ($E_{PL}$) as a function of *x* is well described by the equation:

$$E_{PL}(x) = (1-x)E_{PL}^{MoS2} + xE_{PL}^{WS2} - bx(1-x)$$

, where *b* is the bowing parameter.[5] Here we find $b = 0.17 \pm 0.01$ eV.

## 5. Linear scaling density functional theory calculations of Mo$_{1-x}$W$_x$S$_2$

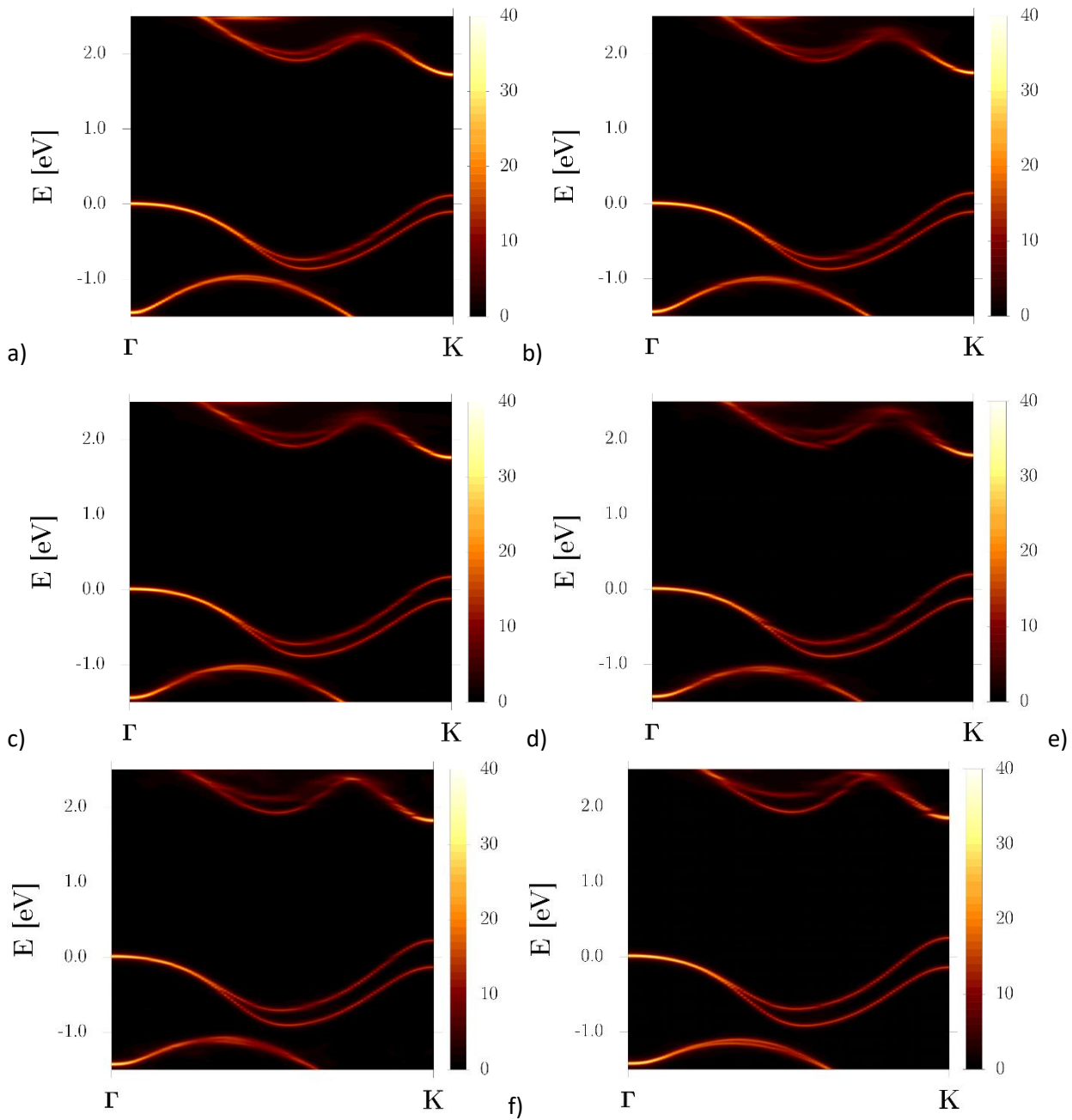

Figure S7: Spectral function intensity plots, derived from LS-DFT calculations for random Mo$_{(1-x)}$W$_x$S$_2$ alloy supercells. Results are projected into the first Brillouin zone of the primitive cell, along the **Γ** to **K** path, for a range of concentrations: a) $x = 0.250$, b) $x = 0.375$, c) $x = 0.500$, d) $x = 0.625$, e) $x = 0.750$, and f) $x = 0.875$. The evolution of the SOC splitting at the VBM can be observed, as can the effect of the disorder potential on the broadening of the conduction band.

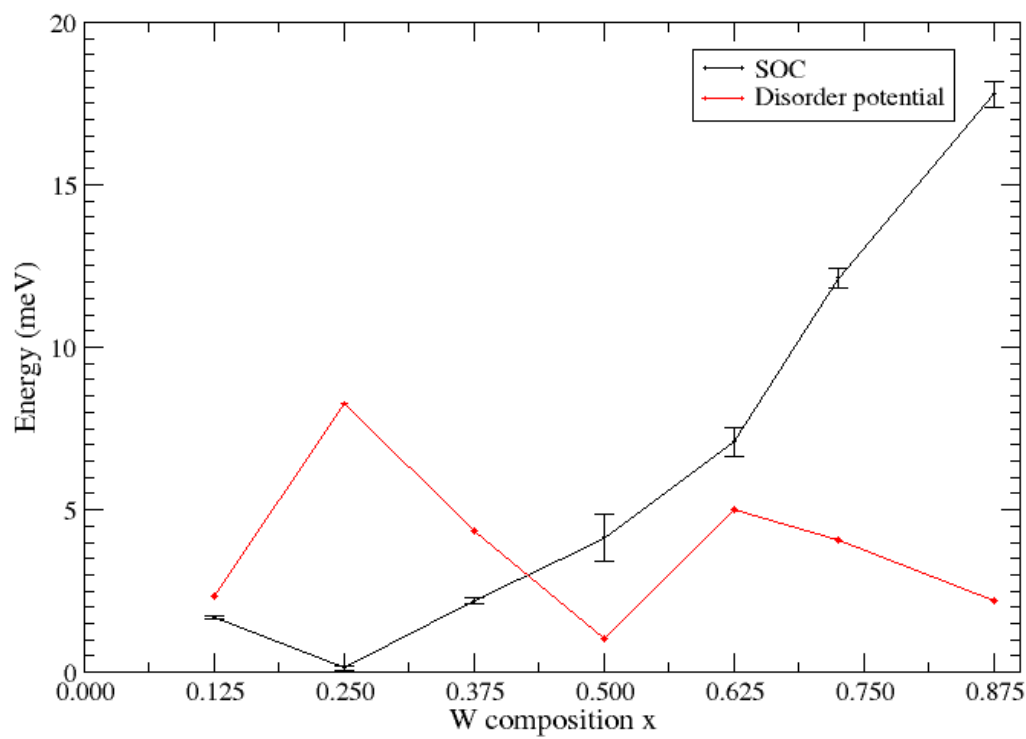

Figure S8: Estimate of the broadening of the conduction band induced by the disorder potential (red line), as measured by the spread of CBM positions across different random realisations of the supercell, compared to the SOC splitting on the conduction band (black). The magnitude of the disorder potential exceeds the SOC for x<0.5.